\documentclass[fleqn,usenatbib]{mnras}
\usepackage[T1]{fontenc}
\DeclareRobustCommand{\VAN}[3]{#2}
\let\VANthebibliography\thebibliography
\def\thebibliography{\DeclareRobustCommand{\VAN}[3]{##3}\VANthebibliography}

\usepackage{graphicx}	
\usepackage{amsmath}	
\usepackage{amssymb}	
\usepackage[
  separate-uncertainty = true,
  multi-part-units = repeat
]{siunitx}
\usepackage{threeparttable}
\usepackage{newtxtext,newtxmath}

\title[Superhumps in BG~Tri]{Superhumps in the cataclysmic variable BG~Tri}

\author[S. Y. Stefanov et al.]{
S. Y. Stefanov,$^{1,2}$
G. Latev,$^{1,3}$
S. Boeva,$^{1}$
M. Moyseev$^{1,2}$
\\
$^{1}$Institute of Astronomy and National Astronomical Observatory, Bulgarian Academy of Sciences, Tsarigradsko Shose 72, BG-1784 Sofia, Bulgaria\\
$^{2}$Department of Astronomy, Sofia University "St. Kliment Ohridski", James Bourchier 5, BG-1164 Sofia, Bulgaria\\
$^{3}$Astronomical Association Sofia, Tsar~Asen~49, BG-1463 Sofia, Bulgaria
}

\date{Accepted XXX. Received YYY; in original form ZZZ}

\pubyear{2022}

\begin{document}
\label{firstpage}
\pagerange{\pageref{firstpage}--\pageref{lastpage}}
\maketitle

\begin{abstract}
We present a detailed photometric study of the bright cataclysmic variable BG~Tri using ground-based observations mainly from Rozhen Observatory, ASAS-SN, TESS, and WASP sky surveys. We report the discovery of a negative superhump with $P_{-sh} = 0.1515(2)$ days and a co-existing superorbital variation with $P=3.94(53)$ days in data from 2019 and 2020. A positive superhump with $P_{+sh} = 0.1727(14)$~days is also discovered in data from 2006. The obtained negative superhump deficit  $\varepsilon_{-}=0.044(1)$ and the positive superhump excess $\varepsilon_+=0.090(9)$ give us an independent photometric evaluation of the mass ratio $(q)$ of the system, which we find to be $q_- = 0.37(2)$ and $q_+ = 0.40(5)$ respectively. We also present a study of the quasi-periodic oscillations (QPOs) and stochastic variability (flickering) in BG~Tri. The light curves show a rich mixture of simultaneously overlapping quasi-periods ranging from 5 to 25 minutes. The multi-color $(UBVRI)$ photometric observations from Rozhen Observatory reveal the typical increase of the flickering amplitudes to the shorter wavelengths. The recently introduced $A_{60}$ amplitude of the flickering light source in all studied photometric bands is systematically lower when the negative superhump is gone in season 2021.
\end{abstract}




\section{Introduction}
\label{sec:Intro}
Cataclysmic variables (CVs) are close binary systems consisting of a Roche-lobe filling secondary, orbiting an accreting white dwarf primary. The secondaries in CVs are usually a low mass main-sequence star (most often a red dwarf, losing matter through the inner Lagrange point $(L_1)$. The transferred matter forms an accretion disc around the white dwarf, which is the primary source of the system's luminosity \citep{Warner_1995,Hellier_2001}. 

Due to the Roche lobe geometry, the secondary star has a teardrop shape with the apex pointing towards the primary. This, in combination with the orbital motion of the system, results in the so-called ellipsoidal modulations (fainter for low-inclination binaries). Detecting these is a direct way to determine the orbital period of the cataclysmic variables \citep{Hellier_2001}. Along with the orbital periodicity, some CVs display periodic changes in brightness with periods close to the orbital period called superhumps. This phenomenon is commonly seen and comprehensively studied in SU UMa stars - subclass of dwarf novae CVs \citep[e.g.,][]{Hirose_Osaki_1990,Kato_2009a, Kato_2017}.
It is believed that a precessing accretion disc can cause such photometric behaviour. Superhumps can either be positive (apsidal - a few percent longer than the orbital period $(P_{orb})$, or negative (nodal) - a few percent shorter. Superhumps can also appear in other sub-classes of CVs. Reports for 15 nova-likes (NLs - a subclass of CVs in the permanently high photometric state) with positive and 11 NLs with negative superhumps can be found in the literature \citep[e.g.,][]{Pavlenko_IR_2020,Ritter_Kolb,Castro_sig_2021,ilkovi4_2021}. The existence of both periodicities has been observed in some systems - TT~Ari~\citep{Belova_TTAri_2013, TTAri_Kraicheva_Stanishev}, AQ~Men~\citep{ilkovi4_2021}, and LS~Cam~\citep{Rawat_LSCam, Stefanov_LSCam}. Positive superhumps are believed to appear when tidal stresses from the secondary affect the disc and cause apsidal precession in the prograde direction \citep[e.g.,][]{Lubow}. Negative superhumps can be explained by a retrograde nodal precession of a tilted accretion disc. The stream impacts the face of the disc and the retrograde precession causes periods shorter than P$_{orb}$ \citep[e.g.,][]{SPH_M_Wood,Montgomery_humps}.

BG~Tri was discovered and classified as a cataclysmic variable by \citet{Khruslov_BGTri}. Although the system is a bright CV (11.9 mag in $V$), it is poorly studied. Other mentions in the literature include a cross-association with X-ray sources by \citet{Cross_Associaton_Xray}. In \citet{Nuv_excess}, BG~Tri is shown to have significant excess in near UV. In 2018 the star dropped in brightness by 2.5 mag in V and was classified as a VY~Scl subclass by \citet{vsnetalert}. $ $ The variables in this subclass of nova-like CVs are predominately in high photometric state except for rare brightness drops up to 6 magnitudes lasting from a few weeks to a few years.

The first comprehensive study of BG~Tri was published by \citet{Hernandez}, where the system is shown to have the spectroscopic characteristics of RW Sex-type and similar nova-like variables. These variables a have high mass transfer rates and bright accretion discs. The authors estimate the inclination angle as $i=25(5)^{\circ}$, the mass of the primary $0.8 M_{\sun}$, the mass of the secondary $0.3M_{\sun}$, and the binary separation - $1.23 R_{\sun}$. The orbital period is measured to be 3.8028(24)~hours. 

Here we present photometric data of BG~Tri from ground-based observations and sky surveys. We report the detection of superhump variability for the first time in BG~Tri. In addition, we analyse the long and short-term variability - superorbital periodicity, quasi-periodic oscillations (QPOs), and flickering.

\section{Observations, data reduction, and photometry}
\label{sec:Observations}

Our first photometric data were gathered during the 2020 summer school of astronomy and astrophysics "Beli Brezi" with a 25~cm~Newton telescope ($FOV = 51' \times 39'$) equipped with a QHY9~CCD ($5.4\mu m$ $3358 \times 2536$ pix) camera and standard Johnson-Cousins $UBVRI$ filters. Later, in the period from August 2020 to November 2021 several observations in the $UBVRI$ bands were carried out at Rozhen National Astronomical Observatory (NAO) in Bulgaria on the 50/70~cm Schmidt telescope ($FOV = 73'\times 73'$) equipped with FLI PL16803 CCD ($9 \mu m$ square $4096 \times 4096$ pix) and the 2.0~m~RCC telescope ($FOV = 10' \times 10'$) with two-channel focal reducer FoReRo2 \citep{FoReRo}, attached to the Ritchey–Chrétien focus and equipped with two ANDOR iKON-L CCDs ($13.5 \mu m$ square $2048 \times 2048$ pix). One additional light curve was obtained on the PWI SDK 17" (43.2 cm corrected Dall-Kirkham) telescope ($FOV = 43' \times 43'$) at AndromedA Observatory - equipped with Alta Apogee U16M CCD ($9 \mu m$ square $4096 \times 4096$ pix). A total of 70 hours of simultaneous and quasi-simultaneous observations in two to five bands were acquired. The majority of observing runs were longer than two hours. A journal of the observations is presented in Table~\ref{tab:observations}.
\begin{table}
	\centering
	\caption {Journal of observtions}
	\label{tab:observations}
	\begin{threeparttable}
	\begin{tabular}{lccr} 
		\hline
		 Date & Length (hr) & Filters & Telescope\\
		\hline
		2020/08/11 & 2.34 & \textit{U B V R I} & 25~cm Newton\\
		2020/08/17 & 3.72 & B V & 25~cm Newton\\
		2020/08/22 & 1.56 & \textit{U B V R I} & 50/70~cm Schmidt\\
		2020/09/22 & 4.99 & \textit{U B V R I} & 50/70~cm Schmidt\\
		2020/09/23 & 5.88 & \textit{U B V R I} & 50/70~cm Schmidt\\
		2020/09/24 & 3.14 & \textit{U B V R I} & 50/70~cm Schmidt\\
		2020/09/25 & 8.34 & \textit{B V} & 50/70~cm Schmidt\\
		2020/10/13 & 1.84 & \textit{U V} & 2.0~m RCC\\
		2020/10/14 & 4.48 & \textit{U V} & 2.0~m RCC\\
		2020/10/14 & 0.74 & \textit{B V R I} & 50/70~cm Schmidt\\
		2020/12/21 & 0.53 & \textit{U V} & 2.0~m RCC\\
		2021/01/18 & 3.61 & \textit{U V} & 2.0~m RCC\\
		2021/01/18 & 2.41 & \textit{B R I} & 50/70~cm Schmidt\\
		\vspace{0.1cm}
		2021/01/20 & 4.25 & \textit{U B V R I} & 50/70~cm Schmidt\\
		2021/10/04 & 1.80 & \textit{U B V R I} & 50/70~cm Schmidt\\
		2021/10/06 & 1.50 & \textit{U B V R I} & 50/70~cm Schmidt\\
		2021/10/30 & 4.11 & \textit{V} & 43.2~cm Dall-Kirkham\\
		2021/11/06 & 8.21 & \textit{V} & 2.0~m RCC\\
		2021/11/07 & 2.45 & \textit{V} & 2.0~m RCC\\
		2021/11/07 & 2.82 & \textit{U B V R I} & 50/70~cm Schmidt\\
		2021/11/08 & 2.82 & \textit{U B V R I} & 50/70~cm Schmidt\\
		\hline
	\end{tabular}
	\end{threeparttable}
\end{table}
\begin{figure}
    \centering
    \includegraphics[width=\columnwidth]{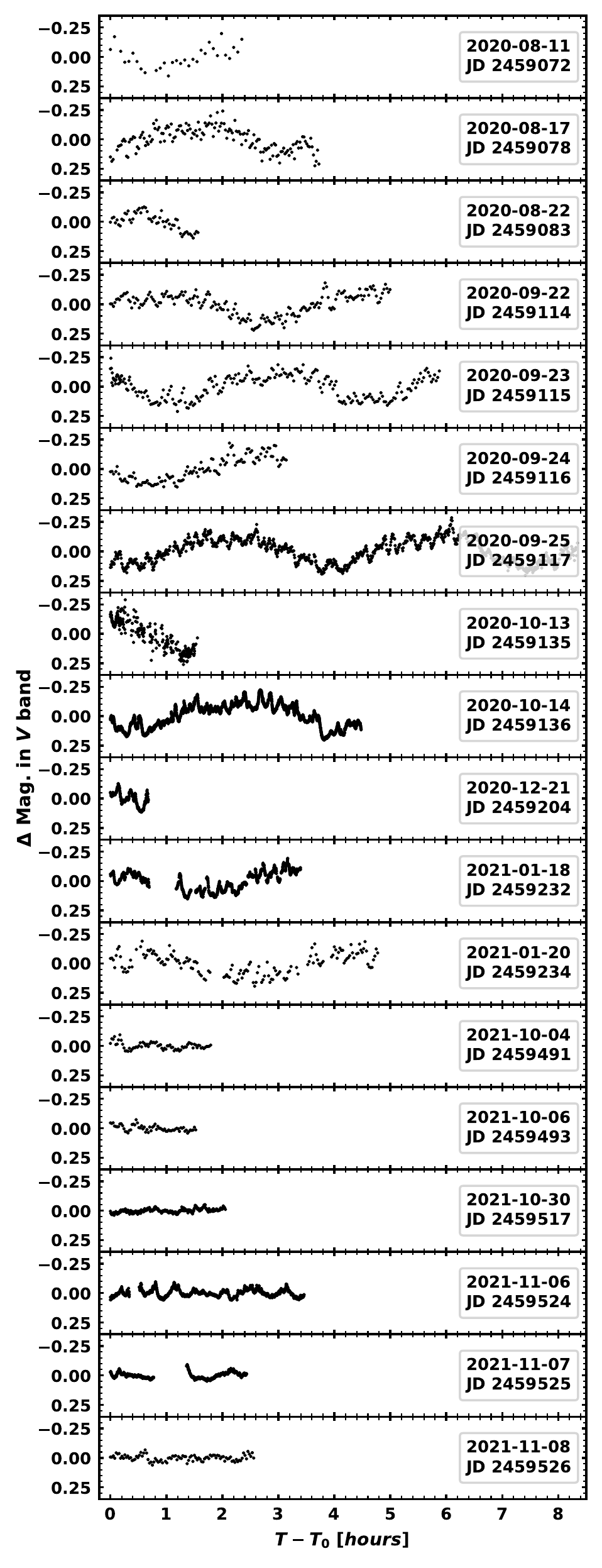}
    \caption{All light curves in V band listed in Table~\ref{tab:observations}. The calendar date and Julian date of each observation are in the right side of the panels. The light curves are shown in relative magnitude (average subtracted) for clarity.} 
 \label{fig:Allv}
\end{figure}
All photometric data were dark or bias subtracted and flat-fielded. To extract the stellar magnitudes, standard aperture photometry was applied using two to five comparison stars depending on the angular size of the frame. The gathered light curves are shown in Figure~\ref{fig:Allv}. 

\begin{figure}
 \includegraphics[width=\columnwidth]{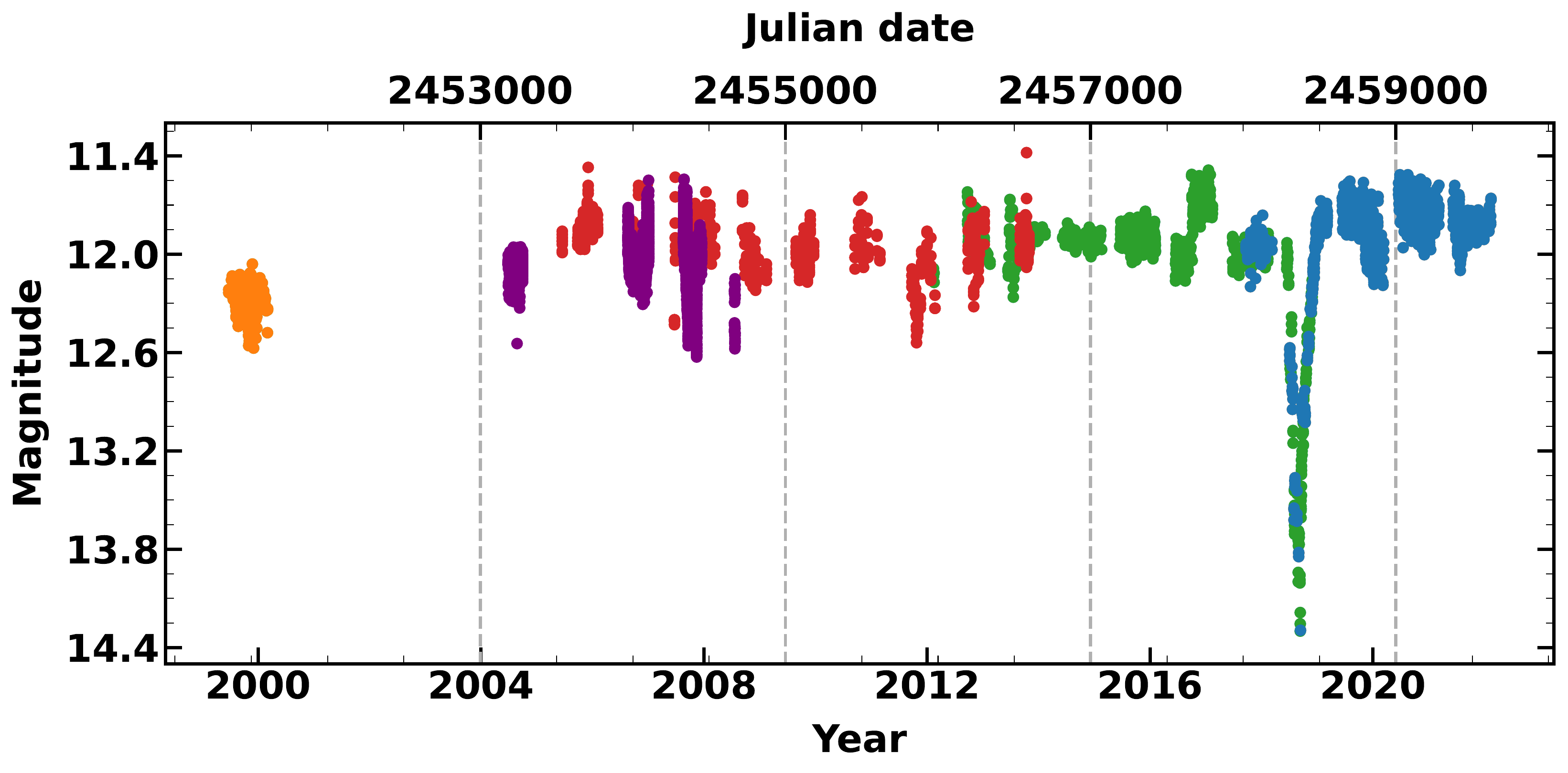}
 \caption{The long term behaviour of BG~Tri in 1999-2021. This light curve is built with data from several sky surveys, but the magnitudes are not transformed to a single band. The color key is as follows  - NSVS magnitudes (orange), WASP magnitudes (purple), CRTS magnitudes (red), ASAS-SN $V$ band magnitudes (green) and ASAS-SN $g$ band magnitudes (blue).}
 \label{fig:Fig1}
\end{figure}

In addition to our photometry, data from several sky surveys was also analysed - The All-Sky Automated Survey for Supernovae~(ASAS-SN)~\citep{ASASSN1,ASASSN2}, Catalina Real-Time Transient Survey (CRTS)~\citep{CRTS}, Wide Angle Search for Planets (WASP)~\citep{WASP}, and the Northern Sky Variability Survey (NSVS)~\citep{NSVS}. A combined light curve is shown in Figure~\ref{fig:Fig1}. 

Photometry of BG~Tri with 1426s cadence from sector 17 of the Transiting Exoplanet Survey Satellite TESS \citep{TESS} was acquired using the Python package "Lightkurve" \citep{Lightkurve} and its dependencies \citep{TESS_cut,astroquery,astropy:2013,astropy:2018}. Custom aperture photometry of the TESS data was performed using this package. The resulting light curve is shown in the top left panel of Figure~\ref{fig:Fig4}.  


\section{Data analysis, results, and discussion}

Sky survey data span 22 years. The mean magnitude of the star in the high state is $\sim$ 11.9~mag in $V$ band and varies with an amplitude of $\sim$~0.5~mag. Only one observed low state was detected in 2018-2019. The decline and rise rates are described in detail in \citet{Hernandez}. The drop in brightness with 2.5 mag is typical for VY~Scl type CVs, but such amplitude shows that the star has not reached a deep minimum of brightness as the ones observed in MV~Lyr, TT~Ari, and KR~Aur \citep[e.g.][]{VYScl_Lowstates}. The data from ASAS-SN and CRTS was converted to V band, joined, and then sent to periodogram analysis. No significant intrinsic large-scale periodicity was found in this data set. 

\subsection{Search for superhumps}
\label{sec:superhumps}
\begin{figure}
 \includegraphics[width=\columnwidth]{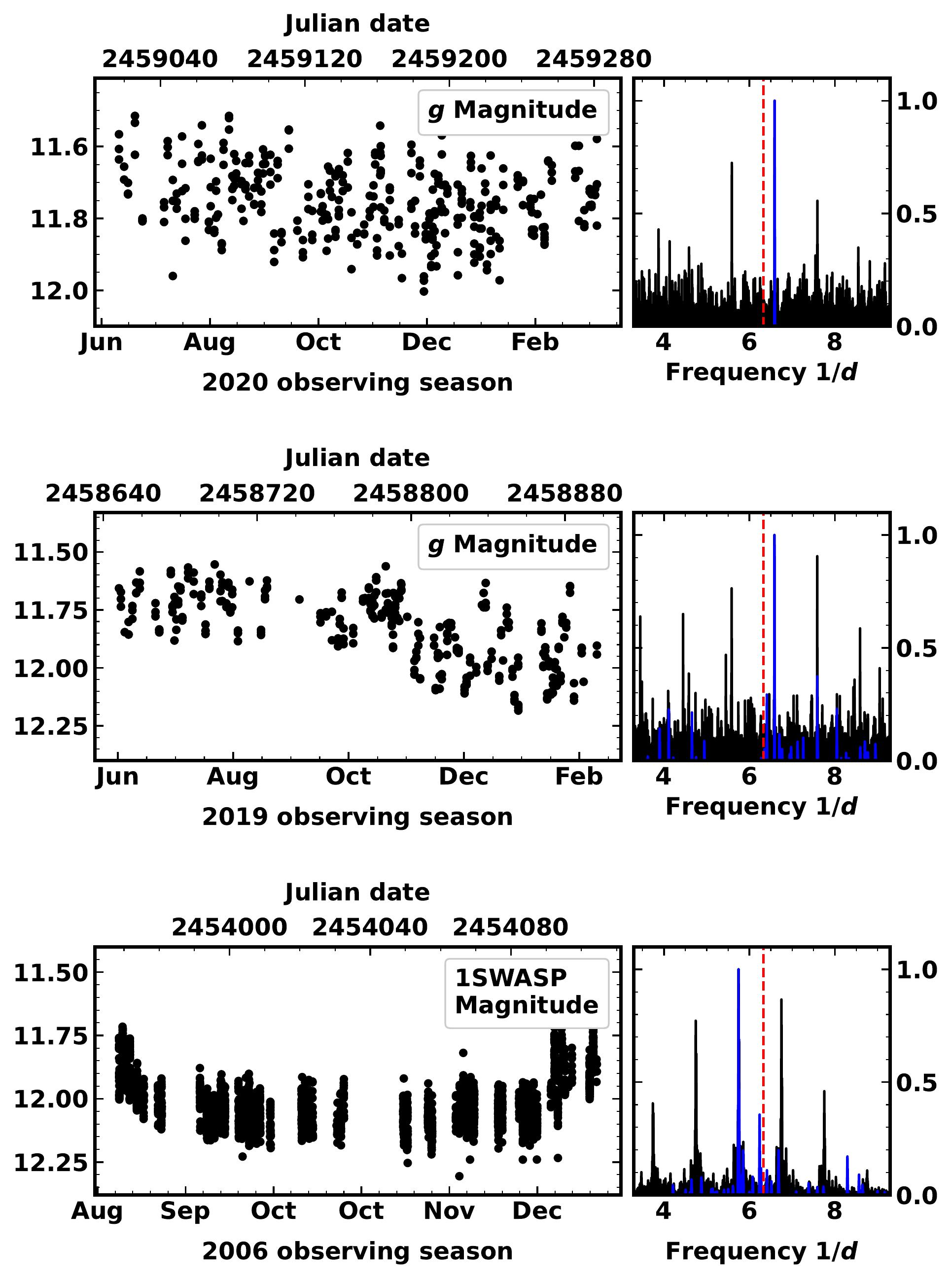}
 \caption{Normalised power spectra of the observing seasons containing significant periodic signals. Light curves and power spectra are shown on the right and left respectively. The top two rows contain data form ASAS-SN. The bottom row shows data from the WASP survey. The CLEAN periodograms are shown in blue and Lomb-Scargle in black. The red dashed vertical line is the orbital frequency of BG~Tri.}
 \label{fig:Fig3}
\end{figure}

Our first two observations of BG~Tri in August 2020 showed light curves with large amplitude variations. The subsequent multi-band monitoring of the star reveals variations in brightness within four hours. The expected ellipsoidal variations would produce double orbital frequency modulation and the secondary shows no significant contribution to the total flux \citep{Hernandez}. Another argument against this interpretation of the variations is that due to the low system inclination, $i=25(5)^{\circ}$, the system is visible face-on, and these would be insufficient to modulate the light curves. Thus, the obvious 4-hour variations may be evidence for the existence of superhumps. To further study this, several long consequent observational runs were carried out at NAO Rozhen. The obtained data are presented in Figure~\ref{fig:Allv}. All data points were analysed using the Lomb-Scargle \citep{Lomb, Scargle} and CLEAN algorithms \citep{CLEAN_algorithm}. A strictly periodic signal with $P=0.1516(2)$~days was detected in 2020 light curves. This period is slightly shorter than the orbital period of the system and represents a negative superhump. The next step was to analyse the available sky survey data. They were separated in seasons of visibility - BG~Tri is visible in the sky from June to February. The observing season is labeled by the year of the beginning of observations, even if some measurements were made in the first months of the next year. The seasons with enough data points were analysed with the same period search algorithms cited above. Strong periodic signals are found in seasons 2006, 2019 and 2020. The resulting power spectra are shown in Figure~\ref{fig:Fig3}. Season 2006 is entirely made of photometry from the WASP sky survey, modulated with $P=0.1727(14)$~days - 9$\%$ longer than the orbital period - a positive superhump. The ASAS-SN photometric data from 2019 and 2020 are in $g$ band. Both seasons contain a periodic variation with $P=0.1515(2)$~days, a negative superhump 5$\%$ shorter than P$_{orb}$. All measured periods and their uncertainties are shown in Table~\ref{tab:hump_periods}. We did not find significant evidence for the presence of superhumps during the observations in 2017 and 2018 of \citet{Hernandez}.
\begin{table}
	\centering
	\caption {Significant periods from the power spectra.}
	\label{tab:hump_periods}
	\begin{tabular}{cl} 
		\hline
		 Data set & Period [days]\\
		\hline
        NAO Rozhen season 2020 & 0.1516(2)\\
        ASAS SN season 2020 & 0.1515(2)\\
        ASAS SN season 2019 & 0.1514(2)\\
        TESS sector 17 2019 & 0.1522(8)\\ 
        TESS sector 17 2019 & 3.94(53)\\
        WASP season 2006 & 0.1727(14)\\
		\hline
	\end{tabular}
\end{table}

\begin{figure}
 \includegraphics[width=\columnwidth]{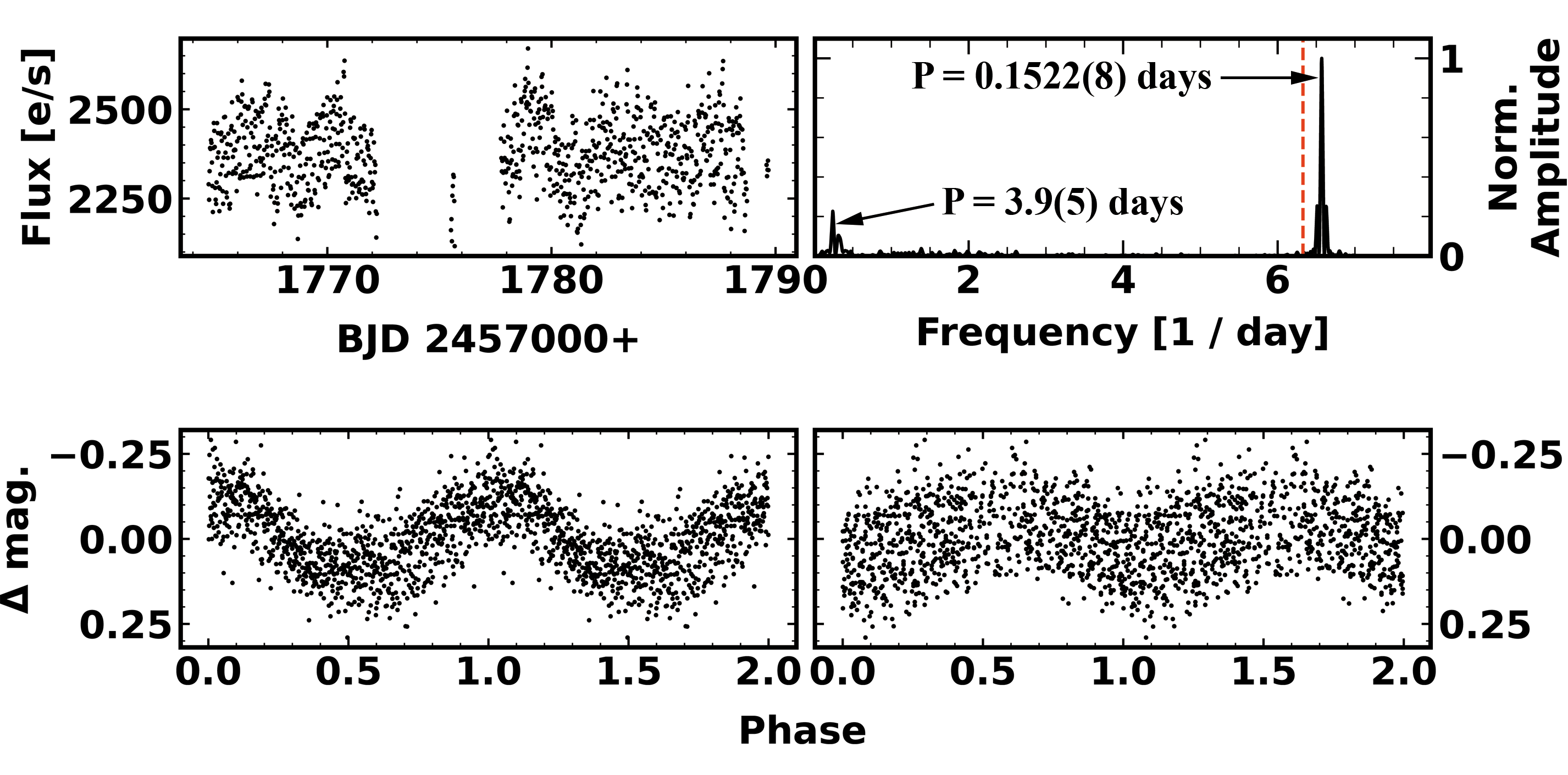}
 \caption{Top: The extracted TESS light curve of BG~Tri and its Lomb-Scargle power spectrum. The red vertical line represents the orbital frequency of the binary. Bottom: Data phased with $P =0.152$~days on the left and with $P =3.94$~days on the right.}
 
 \label{fig:Fig4}
\end{figure}

TESS observations provide continuous photometry of BG~Tri from 07.10.2019 to 02.11.2019, with two gaps - one due to data transfer during perigee and one due to low-quality data. The periodogram analysis shown in Figure~\ref{fig:Fig4} reveals two significant periodicities:
\begin{enumerate}
    \item Negative superhump with $P =0.1522(8)$~days. The superhump period from ASAS-SN data of the same season has a matching period within the uncertainty range.  
    \item Superorbital variation with $P =3.94(53)$~days. This variation is not present in ASAS-SN data, likely due to the limited amount of data and low time resolution.
\end{enumerate}

\subsection{The negative superhumps of BG~Tri}
Negative superhumps are interpreted as a result of the gas stream sweeping across the face of a tilted disc. In a non-tilted disc, the disc overflow is the same on both sides, but in a slightly tilted disc, the overflow would be bigger on one side. If the disc is precessing, the overflow would then be modulated with the beat period of the disc precession and the orbital period of the binary, thus, creating the negative superhump. If this is the case, we would expect the precession period of the disc $P_{prec}$ to be:
\begin{equation}
    \dfrac{1}{P_{prec}} = \dfrac{1}{P_{-sh}} - \dfrac{1}{P_{orb}}.
    \label{eq.beat}
\end{equation}
Using the negative superhump period from TESS sector 17 ($P_{-sh} =0.1522(8)$~d) and the orbital period $P_{orb} = 0.15845(10)$~d from \citet{Hernandez}, the expected value for $P_{prec}$ is $3.8(5)$ days. The other major periodicity found in the TESS data is $P= 3.94(53)$~days, co-existing with the negative superhump period. This we interpret as the precession period of the tilted accretion disc. The precession phase curve is on the bottom right panel of Figure~\ref{fig:Fig4}. The simultaneous presence of negative superhumps and superorbital period in TESS the data is good evidence that both result from a tilted precessing disc.

\begin{figure}
\centering
\includegraphics[width=0.96\columnwidth]{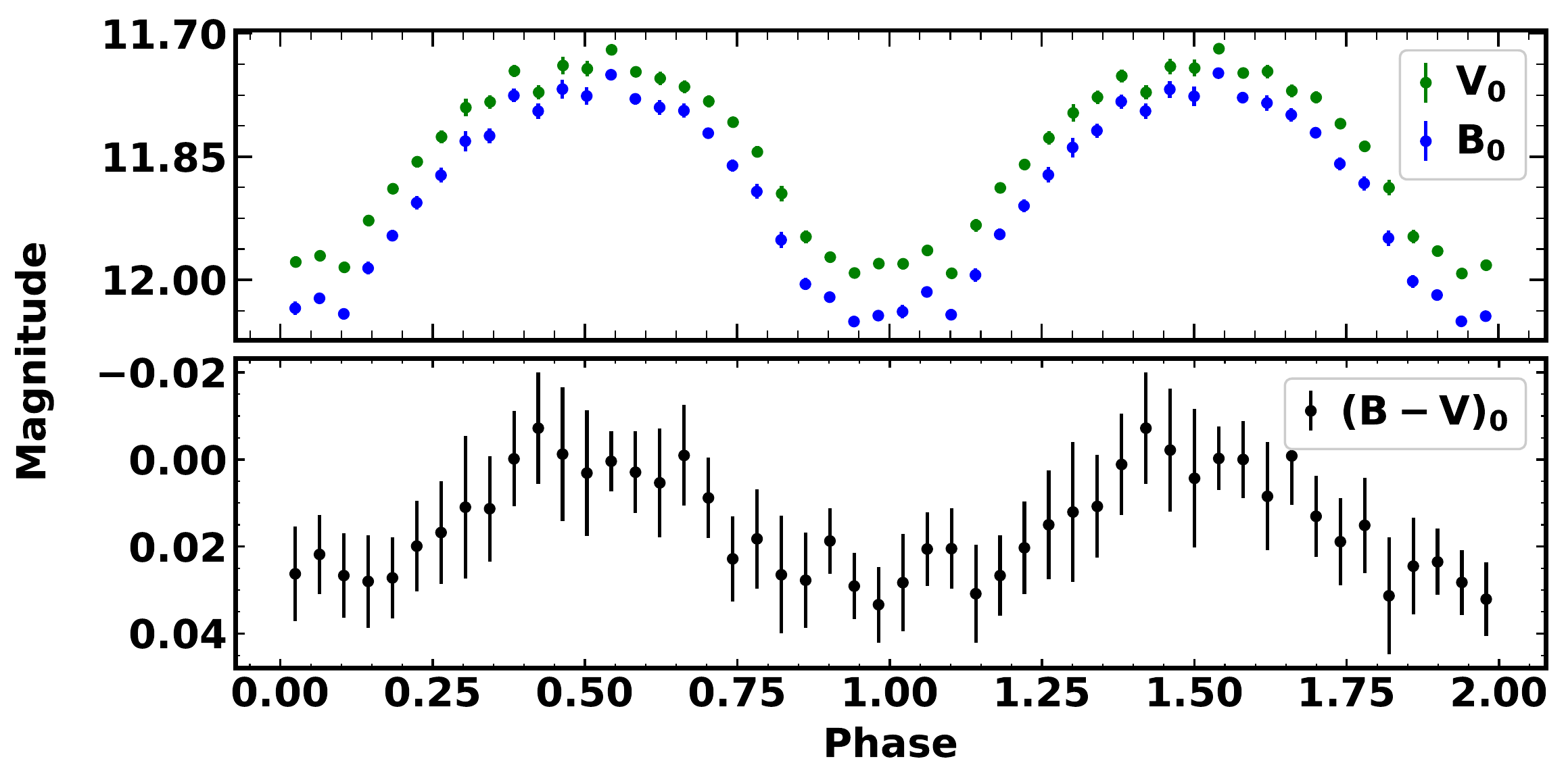}
\caption{The upper panel shows the binned phase curve (25 bins) of the negative superhump in the dereddened $B$ with blue dots and in dereddened $V$ with green dots. The error bars represent the standard error of the data in each bin. The bottom panel is the resulting $(B-V)_0$ color for each bin.}
 \label{fig:Fig7}
\end{figure}

Our multi-color observations show the full amplitudes of the negative superhump in the five bands to be - $\Delta U = 0.30$ mag, $\Delta B = 0.28$, $\Delta V = 0.24$, $\Delta R = 0.22$, and $\Delta I = 0.20$ mag. The $B$ and $V$ magnitudes of the used comparison stars are taken from \citet{APASS}. To account for the interstellar extinction, we used $E(B-V) = 0.03$ from \citet{extincion}. The $(B-V)_0$ color of BG~Tri phased with the negative superhump period is shown in Figure~\ref{fig:Fig7}. This $(B-V)_0$ phase curve is created using phase-binned photometry in the two bands from multiple nights with an equal cadence. Its bluest peak corresponds to the maximum brightness of the superhump. This result is similar to the color index behaviour of ER~UMa \citep{Imada_superhumpcolors} and differs from the case of MASTER~J1727, where the bluest peak coincides with the ascending branch of the phase curve \citep{Pavlenko_prezentaciya}. 


\subsection{Disc tilt}
The tilt angle of the disc is a function of the phase curve amplitude of the nodal precession variability. As the disc precesses, its projected area increases and decreases periodically, thus creating the superorbital variability. Most systems that exhibit this type of disc precession have tilts of a few degrees \citep[e.g.,][]{Kimura_KIC_1,Kimura_KIC}. BG~Tri's disc tilt can be estimated using the amplitude-tilt relation derived in \citet{Smak_2009_humps}:
\begin{equation}
   \delta = A*52.770 \dfrac{(1- u + u\cos{i})\cos{i}}{(1- u + 2u\cos{i})\sin{i}},
    \label{eq.tilt}
\end{equation}
where $u$ is the limb darkening coefficient, $i$ is the system inclination, and $A$ is the semi-amplitude in magnitudes. Using $u=0.6$, $i=25^{\circ}$ from \citet{Hernandez}, and $A = 0.05$ mag, measured by fitting a sine curve to the phase plot of the precession variability (bottom right panel on Fig.~\ref{fig:Fig4}), a value of $\delta \approx 3^\circ$ is obtained. This is the typical value for disc tilts in CVs that display negative superhumps \citep{Smak_2009_humps}.

\subsection{The positive superhumps of BG~Tri}
\begin{figure}
\centering
\includegraphics[width=0.8\columnwidth]{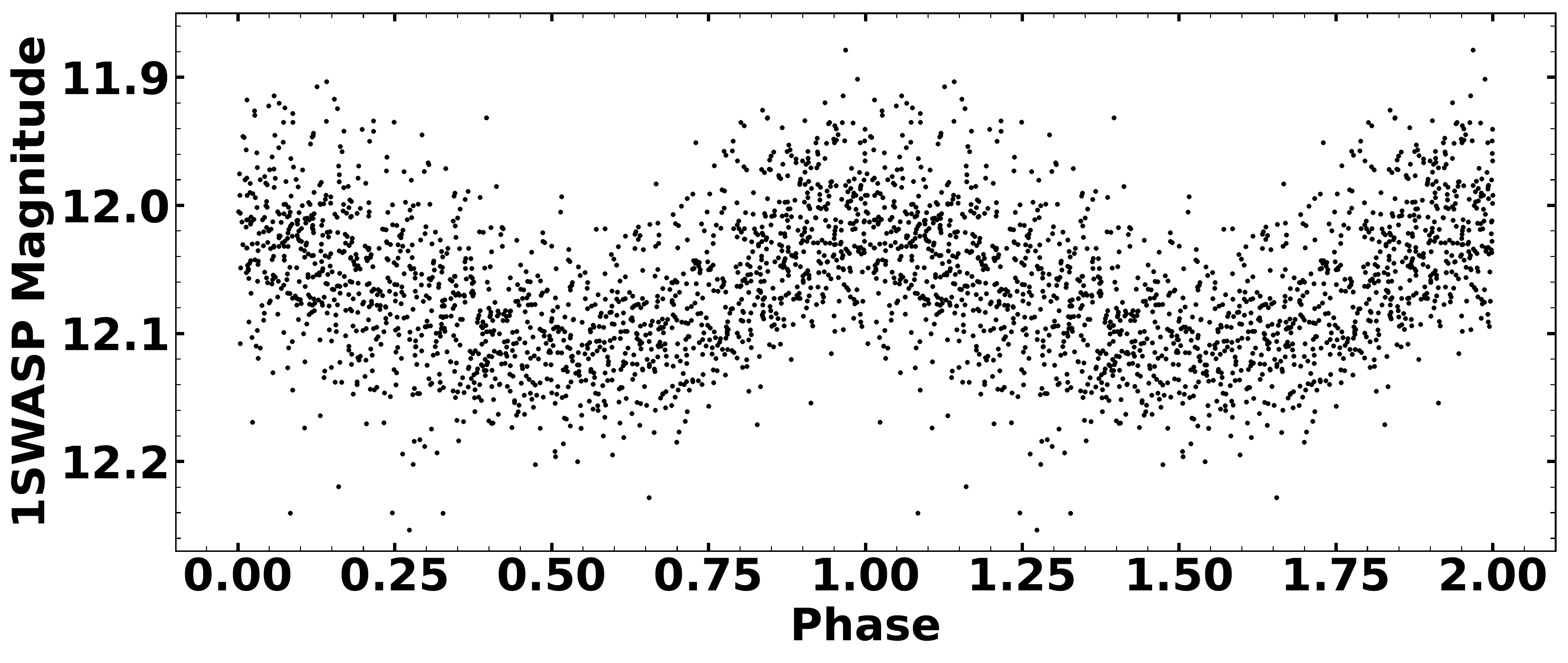}
\caption{Phase curve of the positive superhump of BG~Tri - the WASP data from 2006 folded with $P = 0.1727$~days.}
 \label{fig:Fig_pos_fold}
\end{figure}

Positive superhumps are interpreted as a result of the apsidal precession of an elliptical accretion disc, driven by a 3:1 resonance \citep{Whitehurst_1988, Hirose_Osaki_1990, Lubow}. The resonance induces tidal deformations in the disc, which manifest as brightness variations due to periodic heating. Positive superhumps in BG~Tri are discovered in archival WASP data from 2006. This is the only dataset with detected positive superhumps. Phase curve of the data folded with the most significant period - $P_{+sh} = 0.1727(14)$~days is shown in Figure~\ref{fig:Fig_pos_fold}.
\subsection{System parameters}
\label{sec:syspar}
Superhumps can be used to reveal the mass ratio $q = M_2 / M_1$ using the excess/deficit of the positive/negative superhump:
\begin{equation}
    \varepsilon_{+/-} =  \dfrac{P_{+/-sh} - P_{orb}}{P_{orb}}.
\end{equation}
For BG~Tri the values of $\varepsilon_-$ and $\varepsilon_+$ are $0.044(1)$ and $0.090(9)$ respectively. The ratio $\varepsilon_+ / \varepsilon_- = 2.0(2)$ is in agreement with the value of $\sim 2$ for systems that show both types of superhumps \citep{Patterson_V603, SPH_M_Wood}. The excess and deficit are shown to correlate with $q$. In \citet{SPH_M_Wood}, relation between $\varepsilon_{-}$ and $q$ is derived from particle simulations:
\begin{equation}
q = -0.192|\varepsilon_{-}|^{1/2} + 10.37|\varepsilon_{-}| - 99.83|\varepsilon_{-}|^{3/2} + 451.1|\varepsilon_{-}|^2.
    \label{eq:negative_eps}
\end{equation}
This relation gives a value of $q = 0.37(2)$. For the positive superhump, we used the experimentally derived relation $q(\varepsilon_+)$ in \citet{Kato_shA_2022}, assuming that the measured $\varepsilon_+$ is a stage B superhump excess. This gives us again $q = 0.40(5)$. The uncertainty estimation for $q$ is done using only errors from the periodogram analysis. These relations provide values for $q$ that are in agreement with \citet{Hernandez}, where the WD mass was assumed to be $0.8M_{\sun}$ since this is the average WD mass for binaries above the period gap \citep{Zorotovic_post_common_envelope}. 
With the average value of the system's mass ratio $q = 0.39(3)$ (averaged by the positive and the negative superhump estimations) and mass of the secondary $0.35 M_{\odot}$\footnote{The same result can be obtained with the mass-radius relation in \citet{Demircan}, using the size} of the secondary $R_2 = 0.40 R_\odot$ by Eq. (2.100) in \citet{Warner_1995}. from the mass-period relationship in \citet{Warner_1995}, we calculate the mass of the white dwarf $M_1=0.91(7) M_{\odot}$. Using these parameters, 
the binary separation is $a = 1.33(2) R_\odot$ and the truncation radius of the accretion disc\footnote{ Given by Eq. (2.61) $r_{d} = 0.60 a /(1+q)$ in \citet{Warner_1995}.}  is $r_d = 0.58(2) R_\odot$. We note that the primary Roche lobe radius derived by \cite{Eggleton_1983} is $r_{L} = 0.62(1) R_\odot$, which means that the accretion disc in BG~Tri fills about 93\% of the primary Roche lobe. This is in good agreement with the maximum size of the discs in non-magnetic nova-like CVs \citep{Harrop_disk_size_1996}.


\subsection{Quasi-periodic oscillations and flickering}
\begin{figure*}
    \centering
    \includegraphics[width=\textwidth]{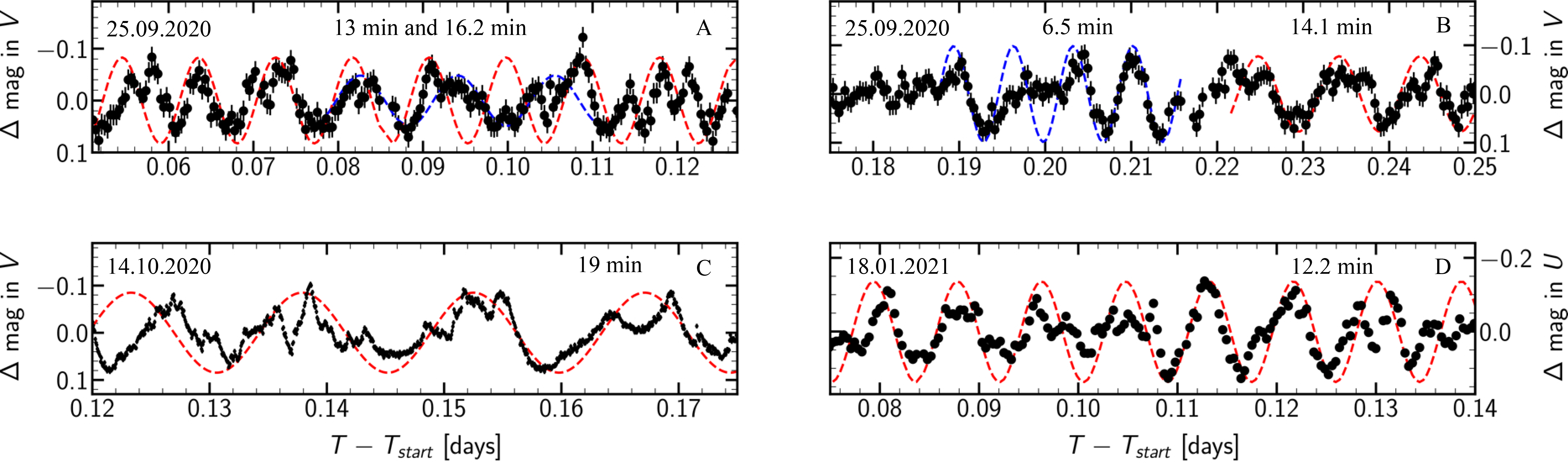}
    \caption{Sinusoidal fits using periods from the spectral analysis of some of the light curves.}

 \label{fig:QPO}
\end{figure*}

\begin{figure}
 \includegraphics[width=\columnwidth]{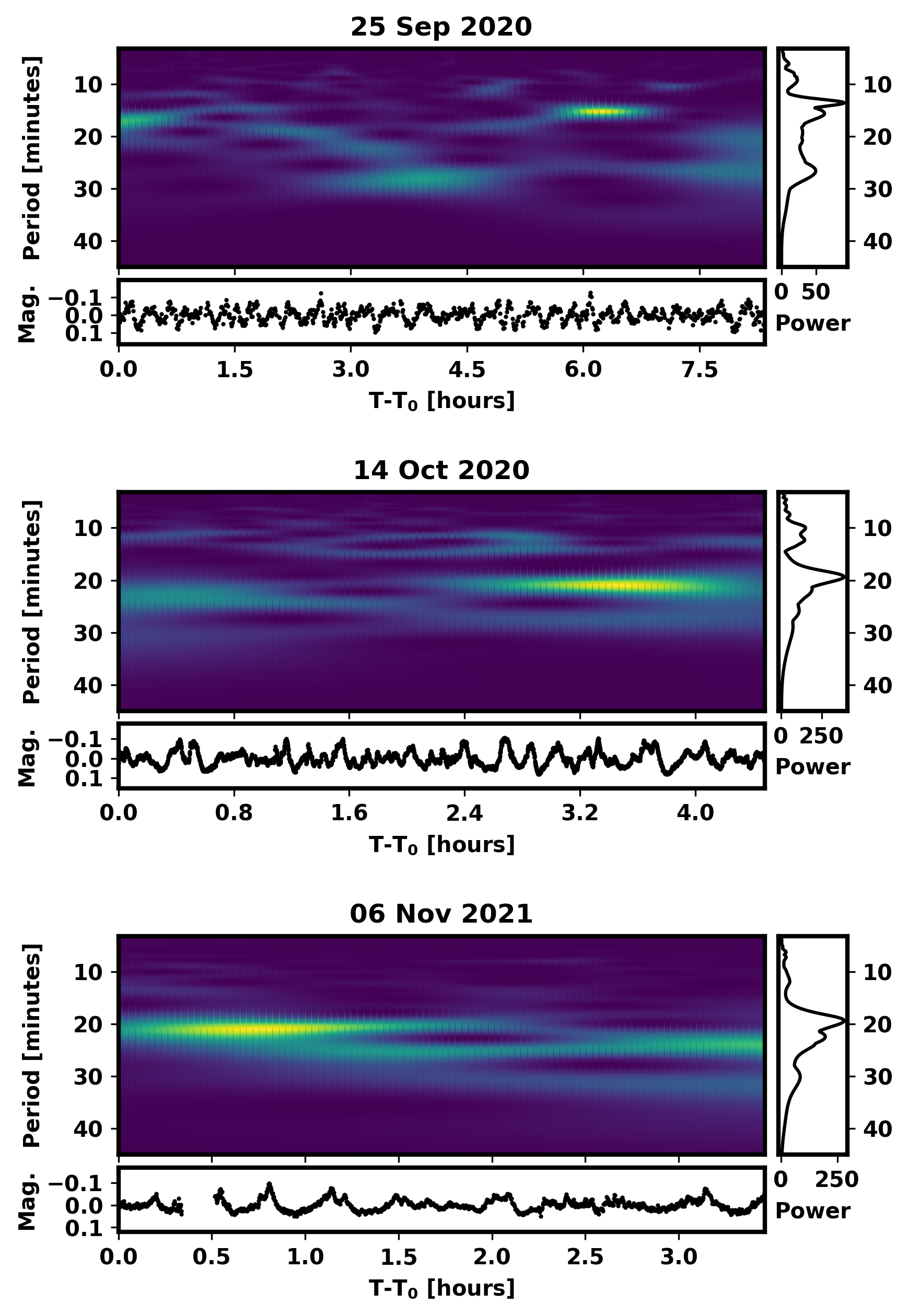}
 \caption{Two-dimensional power spectra of three nights using WWZ.}
 \label{fig:Fig5}
\end{figure}

\begin{figure}
    \centering
    \includegraphics[width=0.85\columnwidth]{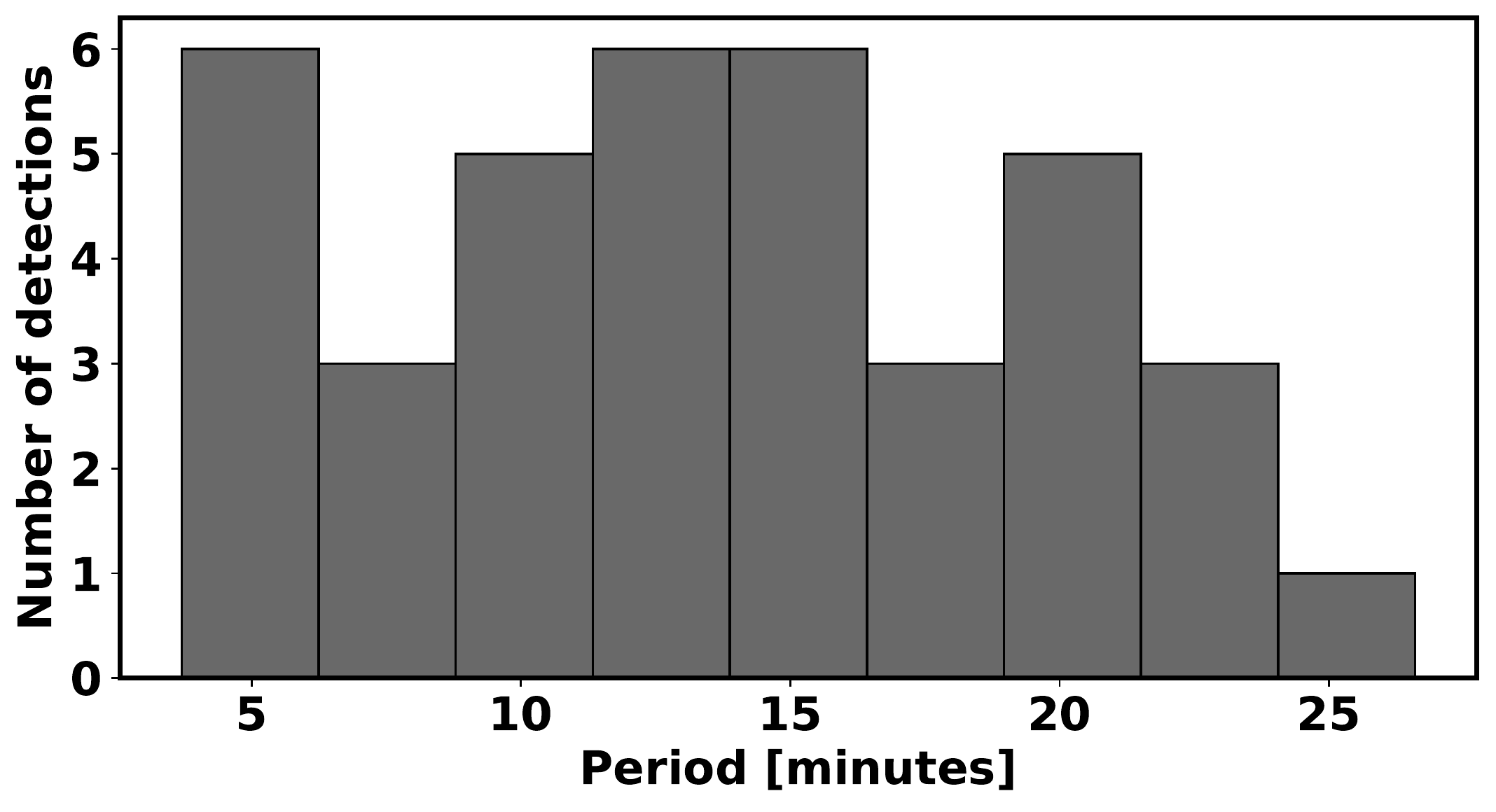}
    \caption{Distribution of the QPO periods found with the WWZ method.}
 \label{fig:A1}
\end{figure}

Fast variations in brightness in timescales of seconds to minutes are typical for cataclysmic variables. Their amplitudes are the largest in $U$ band and can vary from a few hundredths to a few tenths of a magnitude. They are erratic and highly unstable and can change significantly from night to night. QPOs are generally observed in systems with high transfer rates and are believed to be originating from the inner accretion disc. 
To study these phenomena and compare BG~Tri with other similar systems, the approach presented by \citet{Bruch_flickering} was followed. A Savitzky-Golay filter \citep{savgol_filter} with a cutoff time scale $\Delta t = 60$ min and a $4^{th}$ degree smoothing polynomial was applied and subtracted from all light curves in the $V$ band longer than 2 hours. Then, the weighted wavelet Z-transform (WWZ) method by \citet{WWZ} was used to detect and display the evolution of the QPOs in BG~Tri.
A sample of the acquired two-dimensional spectra are shown in Figure~\ref{fig:Fig5}. Typical QPO periods for BG~Tri are in the range of 5-25 min, similar to other CVs - e.g., TT~Ari and V729~Sgr \citep{40yrs_TTAri,V729Sgr_Sun}. A summary of the WWZ analysis for all selected nights is shown in Table~\ref{tab:A1}. Distribution of the detected QPO periods is shown in Figure~\ref{fig:A1}. Overlapping of multiple oscillations with different periods is also present in some of the data. Most of them remain coherent for 3 to more than 10 cycles. Their period and stability vary from night to night - there were no QPOs lasting for more than 3 cycles seen on 14~Oct~2020, but the 14 minute QPO seen in the light curve from 25~Sep~2020 lasts for 11 cycles. Figure~\ref{fig:QPO} shows this behaviour with fitted sines to the QPOs in the data. Sudden changes in period similar to the ones observed in TT~Ari by \citet{TTAri_Kraicheva_Stanishev} are also displayed by BG~Tri. 

If we suppose that the QPOs in BG~Tri occur in dynamical time scale $(t_{dyn})$ related to the Keplerian rotation around the primary, we could determine the possible locations of the observed variations. For $t_{dyn}$ we use the scaled relation in \citet{Sokolski_2003}:
\begin{equation}
\label{eq.tdyn}
    t_{dyn} = 4s \left( \dfrac{r}{10^9 cm} \right)^{3/2}  \left( \dfrac{M_{WD}}{0.6 M_\odot} \right)^{-1/2}.   
\end{equation}
Here $r$ is the distance from the primary (white dwarf) and $M_{WD}$ is the mass of the primary. Replacing the values of $M_{WD}$ and $r_d$ from Section~\ref{sec:syspar} into Eq.~\ref{eq.tdyn} and taking into account their uncertainties, for $t_{dyn}$ we obtain: $t_{dyn} = 14(1)$~min. This result makes it possible to assume that the observed QPOs in the range of $12-14$ minutes are connected with the outer disc rim, or the overflow of matter from the back side of the disc in BG~Tri proposed by \citet{Hernandez}. The full range of the detected QPOs ($5-25 $min.) corresponds to the range of distances $R= 0.28(1) - 0.82(3) R_\odot$ from the white dwarf. The origin of the fastest ones may be related to the formation of 'clumps' or turbulent eddies in the process of transporting matter through the disc.
\begin{table*}
	\centering
	\caption {Average values of the flickering amplitude $A_{60}$ measured in the $UBVRI$ bands for BG~Tri. The numbers in brackets are the standard deviation of $A_{60}$ derived from different light curves and the number of light
curves upon which the average values are based.}
	\label{tab:A60}
	\begin{tabular}{cccccc} 
		\hline
		 Data set & $A_{60}(U)$ & $A_{60}(B)$ & $A_{60}(V)$ & $A_{60}(R)$ & $A_{60}(I)$\\
		\hline
        Season 2020 & 0.13 (0.02) (7) & 0.10 (0.02) (8) & 0.09 (0.02) (8) & 0.09 (0.01) (6) & 0.08 (0.01) (6)\\
        Season 2021 & 0.06 (0.01) (3) & 0.041 (0.004) (3) & 0.04 (0.01) (6) & 0.052 (0.008) (3) & 0.050 (0.008) (3)\\
		\hline
	\end{tabular}
\end{table*}

\begin{table}
	\centering
	\caption {Summary of the quasi-periodic oscillations detected in BG~Tri.}
	\label{tab:QPO_periods}
	\begin{tabular}{cr|cr} 
		\hline
		Date & P [minutes] & Date & P [minutes]\\	
		 \hline
        2020/09/22 & 17.3($\pm$7)    &2021/01/18 & 6.5 ($\pm$1.5) \\
                   & 15.8($\pm$7)    &           & 12  ($\pm$1)   \\
                   & 14.5($\pm$7)    &           & 9.5 ($\pm$1.1) \\
                   & 12.8($\pm$9)    &           & 14.3($\pm$1.3) \\
                   & 10.7($\pm$2)    &           & 4.4 ($\pm$0.6) \\
        2020/09/23 & 18  ($\pm$1)    &2021/10/30 & 20.4($\pm$4.6) \\
                   & 16  ($\pm$1)    &           & 13.3($\pm$1.1) \\
                   & 23.5($\pm$2.5)  &           & 3.7 ($\pm$0.6) \\
        2020/09/24 & 21.5($\pm$3.5)  &2021/11/06 & 19.4($\pm$1.3) \\
                   & 10.1($\pm$1.6)  &           & 22.5($\pm$1.2) \\
                   & 6.8 ($\pm$1.3)  &           & 11.9($\pm$2.2) \\
        2020/09/25 & 13.5($\pm$0.9)  &           & 7.2 ($\pm$0.7) \\
                   & 15.6($\pm$1.6)  &           & 6.2 ($\pm$0.6) \\
                   & 26.6($\pm$2.6)  &           & 5.2 ($\pm$0.3) \\
                   & 19.2($\pm$1.4)  &2021/11/07 & 16.2($\pm$2.7) \\
                   & 6.1 ($\pm$0.8)  &           & 23.9($\pm$3.6) \\
        2020/10/14 & 19.3($\pm$3.4)  &           & 10.2($\pm$2)   \\
                   & 10  ($\pm$1)    &2021/11/08 & 18.4($\pm$1.1) \\
                   & 12  ($\pm$1)    &           & 5.1 ($\pm$0.9) \\
		\hline
	\end{tabular}
	\label{tab:A1}
\end{table}

In his study of flickering in CVs,  \citet{Bruch_flickering} defines the $A_{60}$ amplitude of flickering. After applying the Savitsky-Golay filter as discussed above, the magnitude distribution of each light curve is fitted with a normal distribution, and its FWHM is the $A_{60}$ amplitude. All light curve data were binned to the same cadence before $A_{60}$ amplitudes were measured. The contribution of the secondary is insignificant \citep{Hernandez} and no correction was applied. A minimum of 25 hours of high cadence photometric data in all five bands were acquired during season 2020 on site observations. BG~Tri was in a negative superhump regime on all of these nights. Our data from this season shows that the $A_{60}$ amplitudes are highest in $U$ band - 0.13 mag. The amplitudes in the other filters are similar but decrease to the red bands with $\Delta R$ and $\Delta I$ being 0.09 and 0.08~mag, respectively. In the following season 2021, our observations consist of minimum 8~hours of high cadence data in the five bands. During this season the negative superhump is not present in the light curves, and the $A_{60}$ amplitudes in all bands are half as low those measured in 2020. The amplitudes during season 2020 are similar to other VY~Scl sub-type stars and during season 2021 resemble those of UX~UMa sub-type \citep{Bruch_flickering}. The measured amplitudes from both seasons are shown in Table~\ref{tab:A60}.

\section{Conclusions}
In this work, we present the first photometric study of the bright cataclysmic variable BG~Triangulum. Our light curve analysis yields the following results:
\begin{itemize}
    \item Positive superhumps are discovered in data from the WASP sky survey. They appear only in 2006 data with period $P_{+sh} = 0.1727(14)$ days. This gives a value of the excess $\varepsilon_+ = 0.090(9)$. Using this excess, the mass ratio $q$ is estimated to be $q_+ = 0.40(5)$. 
    
    \item Negative superhumps are discovered in our observations from season 2020 and in data from the ASAS-SN sky survey. They appear in season 2019 after an year-long low VY-Scl state and disappear in season 2021. During these two years, we did not find a significant change in the superhump period - $P_{-sh} = 0.1515(2)$ days. This gives a superhump deficit $\varepsilon_- =  0.044(1)$ and $q_- = 0.37(2)$. 
    
    \item A superorbital variation is present in photometry from the TESS mission. It has a period of $3.94(53)$~days and an amplitude of 0.05~mag. Using this amplitude, we estimate the tilt of the disc of BG~Tri to be $\approx 3^\circ$.
    
    \item Our study of the quasi-periodic oscillations of BG~Tri shows that the
    most common quasi-periods are in the range of $5-25$~min. During our observations in season 2020, BG~Tri was in a negative superhump regime, and we find the amplitude to be the highest in $U$ band $(\Delta U = 0.13$~mag) and decrease to the red bands $(\Delta I = 0.08$~mag). In season 2021 when the superhump is gone, the amplitudes of the QPOs are systematically lower  $(\Delta U = 0.06$~mag, and in $I$ band is $0.05$~mag).
    
\end{itemize}
BG~Tri is a good candidate for the study of CV accretion discs. Its relative brightness in high state allows it to be a great target even for amateur astronomers. The system should be monitored in the future to study the nature of superhump appearance and disappearance.


\section*{Acknowledgements}
We are grateful to the organizers of the "Beli Brezi" summer school of astronomy and the team working with the 25 cm Newton telescope. We also wish to thank the team of AndromedA Observatory for the data they kindly provided us. This study is using publicly available data from the TESS~mission, ASAS-SN, WASP, CRTS, and NSVS. We express our gratitude to the teams of these sky surveys for making their data public.

This study is supported by the grant K$\Pi$-6-H28/2 - Binary stars with compact object (Bulgarian national science fund).

We appreciate the anonymous reviewer's valuable comments and suggestions, which helped us to improve the quality of the manuscript.

\section*{Data availability}
The non-public data underlying this article will be shared on reasonable request to the corresponding authors.



\bibliographystyle{mnras}
\bibliography{bibliography}




\bsp	
\label{lastpage}
\end{document}